\documentclass[aps,preprintnumbers,superscriptaddress,twocolumn,amssymb,floatfix,prl]{revtex4} 
\usepackage{url}
\usepackage[utf8]{inputenc}
\usepackage[T1]{fontenc}
\usepackage[german,english]{babel}
\usepackage{amsmath}
\usepackage{mathtools}
\usepackage{amssymb}
\usepackage{amsthm}
\usepackage{microtype} 
\usepackage{quoting} 
\quotingsetup{font=small}
\usepackage[pdftex]{graphicx}
\usepackage{fancyhdr}
\usepackage[usenames]{color}
\usepackage{bm}
\usepackage{comment}
\usepackage{booktabs}
\usepackage{eurosym} 
\usepackage{braket}
\usepackage{bbold} 
\usepackage{emptypage}
\usepackage{epstopdf}
\usepackage{verbatim}
\usepackage{enumitem}
\usepackage{float} 
\usepackage{chemformula}
\usepackage[normalem]{ulem}

\definecolor{islamicgreen}{rgb}{0.0, 0.56, 0.0}

\newcommand{\be}{\begin{equation}}
\newcommand{\ee}{\end{equation}}
\newcommand{\HH}{\hat{H}}
\newcommand{\At}{\tilde{A}}
\newcommand{\aal}{|\alpha\rangle\langle\alpha|}
\newcommand{\aaa}{|\alpha+1\rangle\langle\alpha|}
\newcommand{\aaadag}{|\alpha\rangle\langle\alpha+1|}
\newcommand{\e}{\textnormal{e}}
\newcommand{\nn}{\hat{n}}
\newcommand{\ii}{\textnormal{i}}
\newcommand{\D}{\hat{D}}

\newcommand{\tn}{\textnormal}

\newcommand{\beq}{\begin{equation}}
\newcommand{\eeq}{\end{equation}}

\renewcommand{\ket}[1]{\left|#1\right\rangle}
\renewcommand{\bra}[1]{\left\langle #1\right|}

\newcommand{\upa}{\uparrow}
\newcommand{\dna}{\downarrow}
{\left\lbrace\begin{array}{@{}l@{}}}%
{\end{array}\right.}

\usepackage[bookmarks=true,colorlinks,linkcolor=blue,urlcolor=blue,citecolor=blue]{hyperref} 

\begin{document}
\title{Vibrational dressing in kinetically constrained Rydberg spin systems}

\author{Paolo P. Mazza}
\affiliation{Institut f\"ur Theoretische Physik, University of T\"ubingen, Auf der Morgenstelle 14, 72076 T\"ubingen, Germany}
\author{Richard Schmidt}

\affiliation{Max-Planck-Institute of Quantum Optics, Hans-Kopfermann-Strasse, 1, 85748 Garching, Germany}
\affiliation{Munich Center for Quantum Science and Technology (MCQST), Schellingstr. 4, 80799 M\"unchen, Germany}
\author{Igor Lesanovsky}
\affiliation{Institut f\"ur Theoretische Physik, University of T\"ubingen, Auf der Morgenstelle 14, 72076 T\"ubingen, Germany}
\affiliation{School of Physics and Astronomy and Centre for the Mathematics and Theoretical Physics of Quantum Non-Equilibrium Systems, The University of Nottingham, Nottingham, NG7 2RD, United Kingdom}
\date{\today}

\begin{abstract}
Quantum spin systems with kinetic constraints have become paradigmatic for exploring collective dynamical behaviour in many-body systems. Here we discuss a facilitated spin system which is inspired by recent progress in the realization of Rydberg quantum simulators. This platform allows to control and investigate the interplay between facilitation dynamics and the coupling of spin degrees of freedom to lattice vibrations. Developing a minimal model, we show that this leads to the formation of polaronic quasiparticle excitations which are formed by many-body spin states dressed by phonons. We investigate in detail the properties of these quasiparticles, such as their dispersion relation, effective mass and the quasiparticle weight. Rydberg lattice quantum simulators are particularly suited for studying this phonon-dressed kinetically constrained dynamics as their exaggerated length scales permit the site-resolved monitoring of spin and phonon degrees of freedom.
\end{abstract}

\maketitle

\emph{Introduction.}-- 
The precise control and manipulation of quantum systems is of utmost importance both in fundamental physics and for applications in quantum technologies. The last decade has seen an immense effort in the improvement of experimental techniques which enable the exploration of quantum many-body systems~\cite{Bloch_rev, Greene2}. Rydberg atoms are notably suitable for this scope due to their versatility in simulating many-body models \cite{Browaeys2020, Greene1, Singer_2005}. In particular, they provide an ideal platform for the realization of spin systems, with applications ranging from quantum information processing~\cite{Quantum-info} to  the exploration of fundamental questions concerning thermalization in quantum mechanics~\cite{Polkovnikov:2010yn, ETHreview, Calabrese_2016}. 

Recently, there has been a growing interest in the study of quantum systems in the presence of kinetic constraints, that impose restrictions on the connectivity between many-body configurations. In particular, it has been observed how constraints, which prevent the system from fully exploring the Hilbert space, can lead to peculiar dynamics and an unexpected lack of thermalization even in systems without explicit symmetries~\cite{Ates2012, Lan2018, Turner2018, Choi2018SU2, Khemani2018, Papic2018long, Pichler2018TDVP, Calabrese_nature, LeroseSurace, LinMot,LinMot2, Shiraishi, AKLT,AKLT2}. A condensed matter manifestation of such effect  can be found e.g. in linear SrCo$_2$V$_2$O$_8$ crystal that is described by a spin-$\frac{1}{2}$ XXZ antiferromagnetic Hamiltonian with a staggered magnetic field~\cite{wang2018experimental}.
\begin{figure}[h!]
\centering
\includegraphics[scale=0.3]{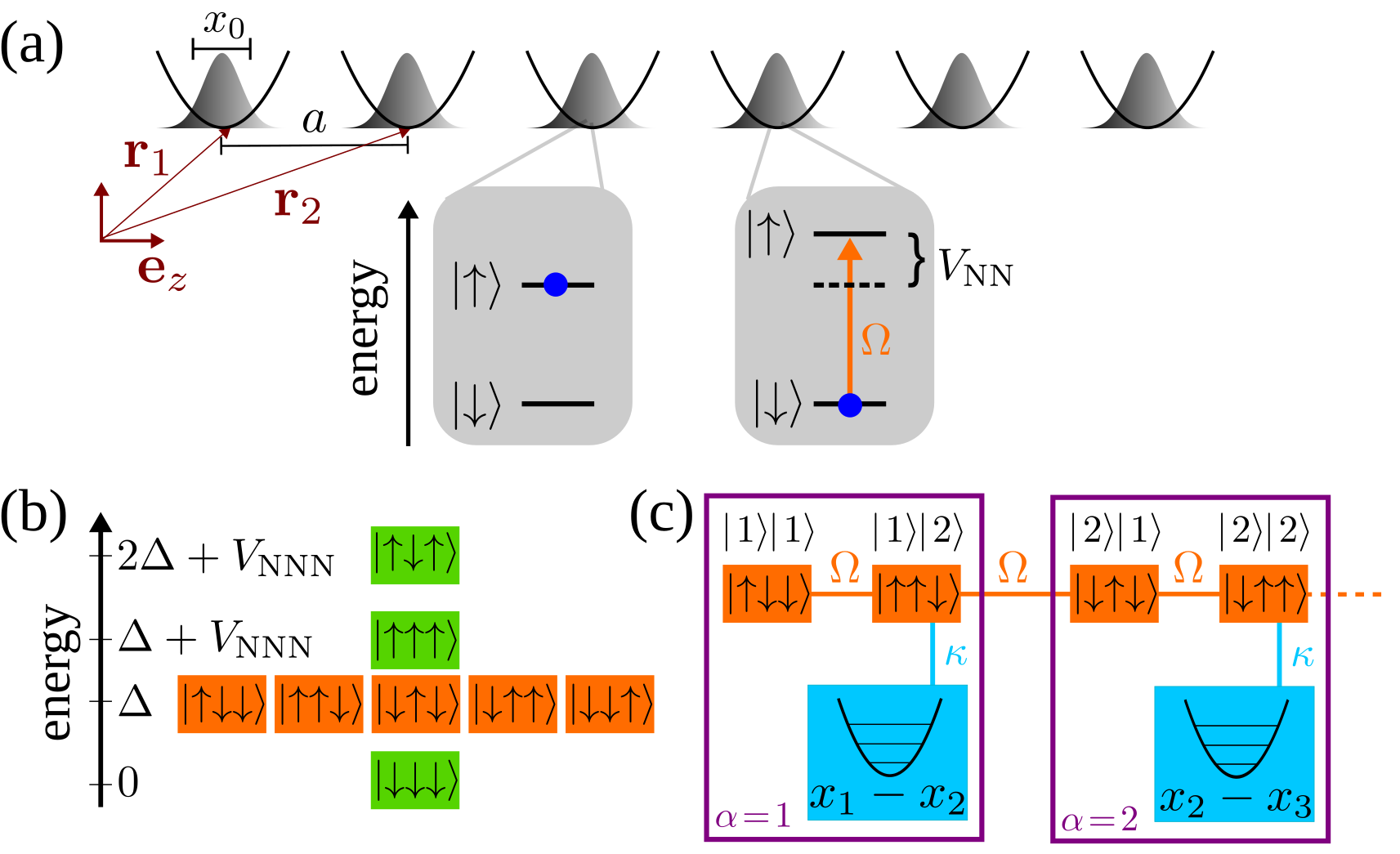}

\caption{\textbf{Setting}: (a) Schematic representation of the Rydberg quantum simulator. The internal degrees of freedom of each trapped atom are approximated by a two-level system. Here, $\Omega$ is the Rabi frequency of the excitation laser that is detuned from the atomic transition by an energy $\Delta$, and the interaction between nearest neighbors is parameterized by $V_\mathrm{NN}$. The atoms are confined in a one-dimensional chain of harmonic traps, where $a$ is the lattice spacing and $x_0$ represents the harmonic oscillator length. (b) Hilbert space representation in the facilitation regime. The  many-body configurations at energy $\Delta$ are resonant and effectively decoupled from the rest of the Hilbert space. $V_\mathrm{NNN}$ is the interaction among next nearest neighbours in Rydberg states. (c) Hilbert space defined by the effective lattice of resonant many-body states and their coupling to harmonic oscillator modes. Only states with two contiguous Rydberg excitations are coupled to the (relative) vibrational mode of the traps.}
\label{Fig:Fig.1}
\end{figure}
Facilitation is a specific instance of a constrained dynamics. The concept was introduced by Fredrickson and Andersen \cite{Fredrickson1984} in the study of kinetic aspects of the glass transition using spin models \cite{Fac3}. Here the excitation of one spin enhances the excitation probability of a neighboring spin. In Rydberg gases such dynamical behavior occurs naturally in the so-called anti-blockade regime \cite{Fac1, Amthor, Fac15, Fac16, Gasenzer_ryd}, and the emerging many-body effects have been investigated in detail in many recent works~\cite{Ostmann_2019,Fac4}. Among the studied phenomena are nucleation and growth  \cite{Schempp2014,Fac2,Urvoy2015,Valado2016,Mattioli_2015}, non-equilibrium phase transitions \cite{Malossi2014,Marcuzzi2016,Letscher2017,Gutierrez2017,Helmrich2020} as well as Anderson \cite{Ostmann_2019,LocFac2} and many-body localization \cite{LocFac}.

In this work we are interested in exploring the interplay between facilitated spin excitations and vibrational degrees of freedom. Such a scenario naturally occurs in Rydberg lattice quantum simulators \cite{Browaeys2020}, where individual atoms are held in oscillator potentials [see Fig.~\ref{Fig:Fig.1}(a)], and coupling between spin and vibrations is caused by state-dependent mechanical forces \cite{Belyansky2019,Gambetta}. We develop a minimal model that describes the emerging complex many-body dynamics and permits a perturbative expansion in the spin-phonon coupling strength. The dressing of the spin dynamics through lattice vibrations leads to the formation of a polaronic quasiparticle \cite{Alexandrov1996} for which we analyse the dispersion relation, the effective mass and the $Z$-factor, determining the quasiparticle weight. The perturbative results are compared with  numerical simulations. Using Rydberg quantum simulators for exploring this physics is particularly appealing as these platforms allow the probing of spin and vibrational degrees of freedom. Thus, using side-band spectroscopy \cite{Kaufman2012}, the phonon cloud that dresses the spin excitation should be directly observable in experiments.

\emph{Facilitated Rydberg lattice.}--
We consider a chain of $N$ traps (e.g. optical tweezers)~\cite{Bernien2017,Barredo:2018aa} each loaded with a single Rydberg atom (see Fig.~\ref{Fig:Fig.1}). The Rydberg atoms can be effectively described as a two-level system in which $|\!\downarrow\rangle_i$ represents an atom in the ground state in the  i$-th$ trap and $|\!\uparrow\rangle_i$ an atom in the Rydberg state. The Hamiltonian of the system is
\begin{equation}
H = \sum_{i=1}^N\left(\frac{\Omega}{2} \hat{\sigma_i^x}+\Delta \hat{n}_i + \sum_{j\neq i}V(\mathbf{r}_i,\mathbf{r}_j)\hat{n}_i\hat{n}_j + \omega a^\dag_ia_i\right),
\label{Eq: Hamiltonian}
\end{equation}
where $i, j$ are indices that label the lattice sites, $\Omega$ is the Rabi frequency, and $\Delta$ is the detuning of the Rydberg excitation laser from the single atom resonance. The interactions among Rydberg states are parameterized by the potential $V(\mathbf{r}_i,\mathbf{r}_j)$ which may be, for example, of  van-der-Waals or dipolar type. Furthermore, we have introduced the spin operators $\hat{\sigma_i}^x=|\!\uparrow\rangle_i\langle\downarrow\!| + |\!\downarrow\rangle_i\langle\uparrow\!|$, $\hat{n}_i= |\!\uparrow\rangle_i\langle\uparrow\!|$. The interaction potential depends on the atomic positions $\mathbf{r}_j = \mathbf{r}^{(0)}_j+\delta \mathbf{r}_j$ where the coordinate of the centre of the $j-$th trap is given by $\mathbf{r}^{(0)}_j = aj\mathbf{e}_z$, with $a$ the lattice constant, c.f.~Fig.~\ref{Fig:Fig.1}. The fluctuations $\delta_j=\frac{\delta r_j}{a}$ around the trap center can be expressed in terms of the bosonic operators (obeying $[a_i, a^\dag_j] = \delta_{i,j}$) as $\delta _j = \frac{x_0}{a}(a^\dag_j + a_j)$, where $x_0=\sqrt{\hbar/(m\omega)}$ is the harmonic oscillator length and $m$ the atomic mass. Assuming that $\delta r_i \ll a$, i.e. the interparticle separation is much larger than the fluctuations around the equilibrium positions, we can expand the interaction potential to first order obtaining a coupling term between the Rydberg excitations and the  vibrational trap modes. Here we are considering only the longitudinal modes because, as shown in the Supplemental material, in one-dimensional lattices the coupling with the transverse modes is negligible at the first order of the perturbative expansion of the potential. In this case we obtain:

\begin{equation}
V(\mathbf{r}_i,\mathbf{r}_j)\simeq V(\mathbf{r}_i^0,\mathbf{r}_j^0) + G(\mathbf{r}_i^0,\mathbf{r}_j^0)(\delta_i -\delta_j).
\end{equation}  
Here, $G(\mathbf{r}_i^0,\mathbf{r}_j^0)$ is the gradient of the potential~\footnote{Note, that these two quantities can be tuned independently, as shown in~\cite{Gambetta}.}.

In the facilitation regime, the interaction between two neighboring atoms is cancelled by the laser detuning, $\Delta  = -V(\mathbf{r}_i^0, \mathbf{r}_{i+1}^0)$. This means that transitions between many-body configurations of the type $|\!\dna\upa\dna\dna\rangle \leftrightarrow |\!\dna\upa\upa\dna\rangle $ become \emph{resonant} [see Fig.~\ref{Fig:Fig.1}(b)]. In order to simplify the dynamics further we assume that the interaction between next-nearest-neighbours is larger than the Rabi frequency, i.e.
$V(\mathbf{r}_i^0, \mathbf{r}_{i+2}^0)\gg |\Omega|$. This prevents the growth of clusters and constrains the evolution of a single initial seed atom to a subspace in which at most two adjacent atoms are excited, e.g. $|\!\upa\dna\dna\dna\rangle \leftrightarrow |\!\upa\upa\dna\dna\rangle \leftrightarrow |\!\dna\upa\dna\dna\rangle \leftrightarrow |\!\dna\upa \upa \dna\rangle \leftrightarrow\ldots$, as shown in Fig.~\ref{Fig:Fig.1}(b). 

This subspace of many-body states defines an effective one-dimensional lattice with a two-site unit cell, for which we introduce the labels $|\alpha\rangle |s\rangle$. Here, the variable $\alpha$ denotes the position of the leftmost excited spin and $s$ the number of excited spins, i.e., for $\alpha=1$,  $|\alpha\rangle | 1\rangle = |\!\upa\dna\dna\dna\rangle$, $|\alpha\rangle | 2\rangle = |\!\upa\upa \dna \dna\rangle $ and $|\alpha+1\rangle | 1\rangle = |\!\dna\upa\dna\dna\rangle$, etc. [see Fig.~\ref{Fig:Fig.1}(c)]. On this effective lattice, the Hamiltonian~\eqref{Eq: Hamiltonian} can be rewritten as
\begin{eqnarray}
H &= &\Omega\sum_{\alpha}\Big(|\alpha\rangle\langle \alpha| \mu^x +\mu^-|\alpha+1\rangle\langle\alpha| +\tn{h.c.}\Big) \nonumber\\
&&+\kappa\sum_{\alpha}\frac{\mu^z-\mathbb{1}}{2}|\alpha\rangle\langle \alpha| (a^\dag_{\alpha+1} + a_{\alpha+1} - a^\dag_\alpha - a_\alpha)\nonumber\\
&&+ \omega\sum_{\alpha} a^\dag_\alpha a_\alpha,
\label{Eq:Ham}
\end{eqnarray}
where $\mu^x=|1\rangle\langle 2|+ |2\rangle\langle 1|$, $\mu^-=|1\rangle\langle 2|$ and $\mu^z= |1\rangle\langle 1|-|2\rangle\langle 2|$ are spin operators, that characterize the two non-equivalent types of sites in the lattice and $\kappa = -\sqrt{\frac{1}{2 m \omega}}G(\mathbf{r}_i^0,\mathbf{r}_j^0)$.\\

\emph{Vibrational dressing.}--
Through Eq.~\eqref{Eq:Ham} it is evident that bosons, which correspond to the trap vibrations, interact only with states where $s=2$, i.e., with states in which there are two adjacent excited spins, as shown in Fig.~\ref{Fig:Fig.1}(c). In order to simplify the description we rewrite the Hamiltonian~\eqref{Eq:Ham} by introducing the Fourier transformed bosonic modes $a_j = \frac{1}{\sqrt{N}}\sum_{p=-N/2}^{N/2}\e^{\ii\frac{2\pi }{N}jp}A_p$, which yields \begin{eqnarray}
H&=&\Omega\sum_{\alpha}|\alpha\rangle\langle\alpha|\mu^x + \Omega\sum_\alpha\big(\mu^-|\alpha +1\rangle\langle\alpha| + \tn{h.c.}\big)\nonumber\\
&&+\frac{\kappa(\mu^z-\mathbb{1})}{2\sqrt{N}}\sum_p \left[\left(\e ^{-\ii\frac{ 2\pi}{N}p}-1\right)\e^{-\ii\frac{ 2\pi}{N}\hat\alpha}A^\dag_p+\textnormal{h.c.}\right]\nonumber\\
&&+\omega\sum_p A^\dag_pA_p.
\label{Eq:Hamiltonian_Fourier}
\end{eqnarray}
where $\hat \alpha = \sum_\alpha \alpha \ket{\alpha}\bra{\alpha}$ denotes the lattice position operator.
We can decouple the lattice from the bosonic modes using the \emph{Lee-Low-Pines} transformation~\cite{Lee-Low-Pines} 

\begin{equation}
U = \exp{\left[-\ii\hat \alpha  \sum_p \frac{2\pi p}{N} A^\dag_pA_p\right]}.
\label{Eq:H_LLP}
\end{equation}
Introducing the Fourier modes of the quasi-particles, $|\alpha\rangle =\frac{1}{\sqrt{N}}\sum_{q=-N/2}^{N/2} \e^{\ii\frac{2\pi }{N}q}|q\rangle$, the transformed Hamiltonian reads $U^\dag H U = \sum_q |q\rangle\langle q| H_q$, with
 \begin{eqnarray}
 H_q &=&\Omega\left[ \mu^+\left(1+
 \e^{-\ii\frac{2\pi}{N}(\sum_p p A^\dag_pA_p +q)}\right) +\textnormal{h.c.}\right]\nonumber \\
 &&+\frac{\kappa(\mu^z-\mathbb{1})}{\sqrt{2N}}\sum_p\left[\left(\e ^{-\ii\frac{ 2\pi }{N}p}-1\right)A^\dag_p+\textnormal{h.c.}\right]\nonumber\\
 &&+\omega\sum_p A^\dag_pA_p.
\label{Eq:Ham_int}
\end{eqnarray}
By virtue of the canonical transformation the quasiparticle momentum $q$ is now a conserved quantum number, which simplifies tremendously the subsequent analysis. Further manipulations, which are detailed in the Supplemental Material, allow us to finally obtain
\begin{eqnarray}
H_q &=&\omega\sum_p \tilde{A}^\dag_p\tilde{A}_p\mathbb{1}-\Omega \cos\left[\frac{\pi}{N}\left(\sum_p p A^\dag_p A_p+q\right)\right]\mu^z\nonumber\\
&&+H^{\tn{I}}_{\tn{int}} + H^{\tn{II}}_{\tn{int}} ,
\label{Eq:Ham_final}
\end{eqnarray}
with $H^{\tn{I}}_{\tn{int}}=-\frac{\kappa^2}{\omega}(\mathbb{1}-\mu^x)$ and the displaced bosonic operators $\tilde{A}_p= A_p +\frac{\kappa}{\omega\sqrt{N}}\left(\e ^{-\ii\frac{2\pi }{N}p}-1\right)$. 
An explicit expression for $H^{\tn{II}}_{\tn{int}}$ is given in the Supplemental Material. Note, that despite the achieved simplification, the Hamiltonian~\eqref{Eq:Ham_final} is highly non-trivial and now describes many-body spin states coupled to a bath of interacting phonons.

\begin{figure}[!t]
\includegraphics[scale=0.45]{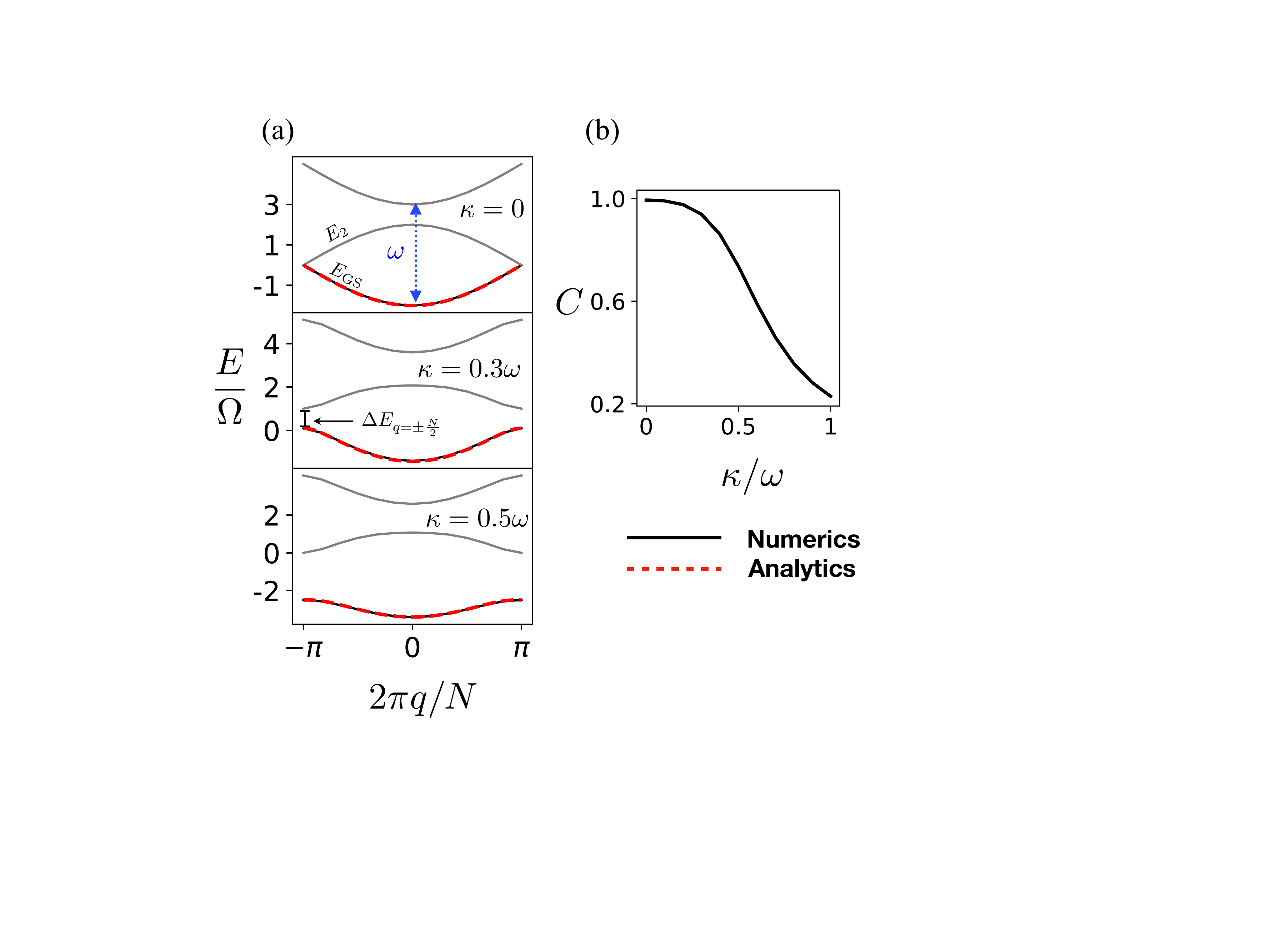}
\caption{\textbf{Band Structure:} (a) Comparison between the analytical form Eq.~\eqref{Eq:correction} and numerical calculations obtained for $N=12$ and truncation of the maximum number of phonons per site to $3$. The non-interacting bands form a ladder with spacing $\omega$. As the interaction between the lattice and phonons increases, the band degeneracies at the edges are lifted and the ground state band is flattened. (b) Curvature $C$ of the ground-state band computed for quasiparticle momentum $q=0$, numerical results.}
\label{Fig:Fig2}
\end{figure}

To investigate the vibrational dressing of the facilitation dynamics we first consider the decoupling limit $\kappa = 0$. In this case the spectrum of Hamiltonian \eqref{Eq:Ham_final} is given by bands that appear in pairs with positive and negative curvature [see Fig. \ref{Fig:Fig2}(a)]. There are infinitely many pairs, forming a ladder with a spacing given by the trap frequency $\omega$. The ground state band has the tight-binding dispersion relation, i.e. $E_{\tn{GS}}=-\Omega\cos\left(\frac{\pi}{N}q\right)$.
Note that in the limit of $N\gg1$, the argument of the cosine becomes a continuous variable $-\pi\leq \frac{2\pi}{N}q \leq \pi$. In the following we assume for simplicity that the trap (phonon) frequency is larger than the twice the laser Rabi frequency, $\omega>2\Omega$. In this case the ground state band is well separated from the remaining ones. Crucially, this regime is within reach of current technology from an experimental point of view. In fact, in order to be able to observe coherent dynamics, we must have $\omega>2\Omega\gg\gamma$ with $\gamma$ being the  decay rate of the Rydberg atoms. Typically, $\gamma\approx10^4~\textnormal{Hz}$ and frequencies larger than $10^5~\textnormal{Hz}$ can be achieved experimentally, for both $\Omega$ \cite{LocFac2} and $\omega$ \cite{Kaufman2012}. Furthermore, both the Rydberg and the ground state ought to be trapped as demonstrated in Ref.~\cite{Barredo}. 

In the presence of interactions between the propagating Rydberg excitation and the phonons, i.e.  for $\kappa\geq 0$, the energy bands, defining the spectrum of Eq.~\eqref{Eq:Ham_final}, are modified. In particular, we observe the lifting of the degeneracy of the ground state and the first excited band at the band edges together with a flattening of the band structure. The decrease of the band curvature, shown in Fig. \ref{Fig:Fig2}(b), is a consequence of the phonon-dressing of the spin excitation which leads to the formation of a polaron quasiparticle which is characterized by a correspondingly increased effective band mass.
\begin{figure*}[!t]
\includegraphics[scale=0.45]{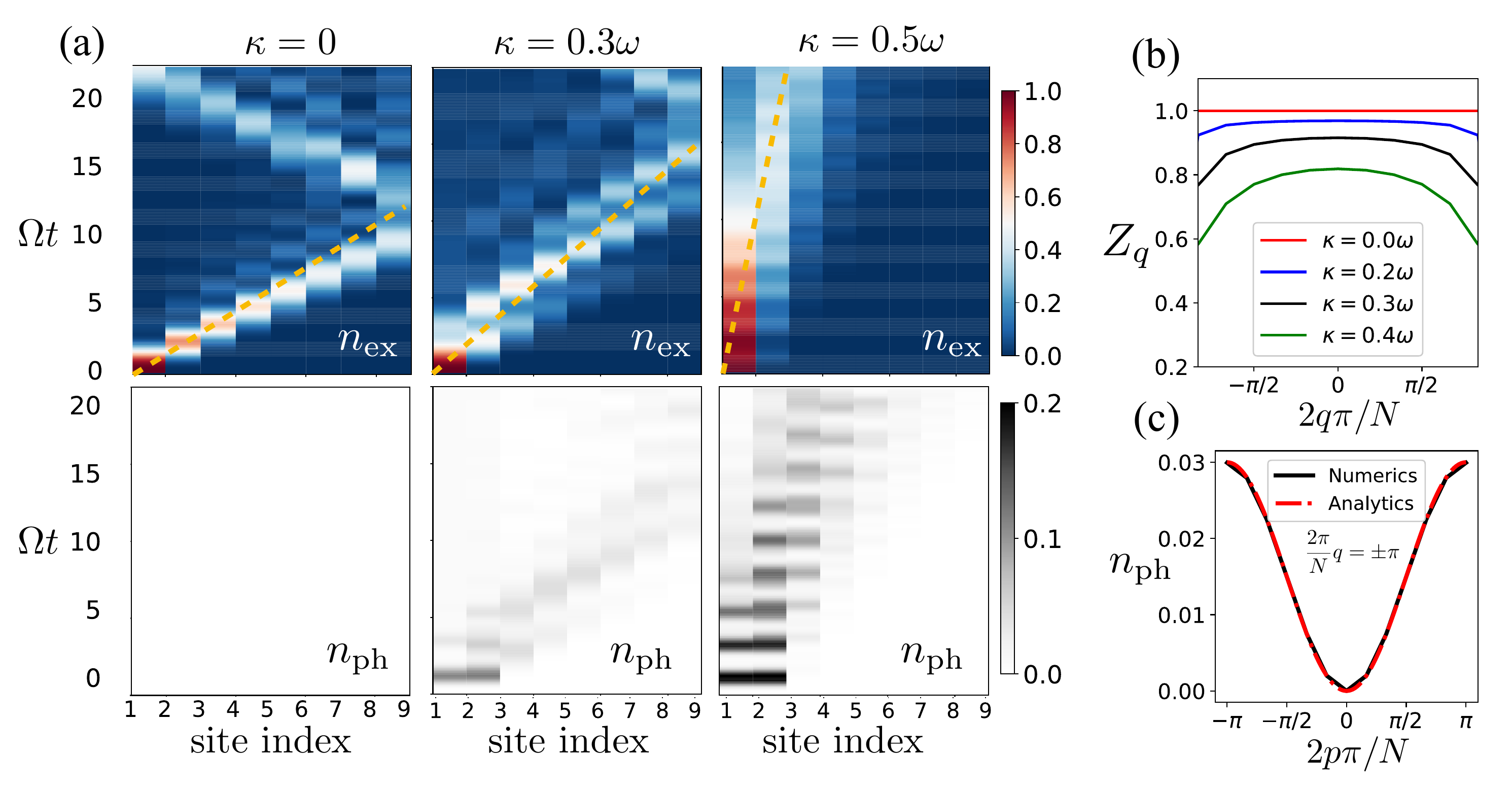}
\caption{\textbf{Polaron dynamics:} (a) Density plot of the Rydberg and phonon excitation for different values of the interaction strength $\kappa/\omega$. The first (second) row shows the Rydberg (phonon) density. For $\kappa=0$, i.e. in absence of interactions, a ballistic spreading of the Rydberg excitation is observed and no phonons are generated (orange dashed lines are a guide to the eye). At $\kappa=0.3\omega$ ($\kappa=1.5\Omega$ and $\omega=5\Omega$) the propagation of the Rydberg excitation is slowed down until it almost comes to a halt (on the timescale shown) when $\kappa=0.5\omega$. (b) Momentum-dependent $Z$-factor for different values of the coupling strength. (c) Occupation number of the phonon modes in momentum space (labelled by $p$) of the ground state for quasi-particle momentum $\frac{2\pi}{N}q=\pm \pi $ and $\kappa=0.3\Omega$. We compare the analytical result of Eq.~\eqref{Eq:numbos1} with numerical data obtained using exact diagonalization on a system up to $N=12$ sites.}
\label{Fig:Fig3}
\end{figure*}
In order to obtain a qualitative understanding of the observed renormalization of the band structure we adopt a perturbative approach to the solution of Eq.~\eqref{Eq:Ham_final}. The term $H^{\tn{I}}_{\tn{int}}$ couples  states with quasiparticle momentum $q$ of the ground state band, $|\Psi^{(0)}_q\rangle$, to the first excited band. At first order in perturbation theory, this correction can be computed solving a two-level eigenvalue problem for each $q$.We have also an additional correction to the energy given by the action of $H^{\tn{II}}_{\tn{int, GS}}$. This yields the \emph{dressed} value for the ground state band, $E^{(1)}_{\tn{GS}}(q)$:
\begin{equation}
E_{\tn{GS}}^{(1)}=-\sqrt{\Omega^2\cos^2\left(\frac{\pi q}{N}\right)+\frac{\kappa^4}{\omega^2}} -\frac{\kappa^2}{\omega} +\frac{\Omega\kappa^2}{\omega^2 N}\xi_N\cos\left(\frac{\pi q}{N}\right),
\label{Eq:correction}
\end{equation}
with $\xi_N = \frac{4 \cos \left(\frac{\pi }{N}\right) \cot \left(\frac{\pi }{2 N}\right) \cos \left(\frac{\pi  q}{N}\right)}{2 \cos \left(\frac{\pi }{N}\right)+1}$. 
As it can be seen in Fig.~\ref{Fig:Fig2}, there is good agreement between the analytical result and the numerics. 

\emph{Dressed facilitation dynamics.}--
The interaction between the Rydberg atoms and the phonons that leads to the phonon dressing and corresponding band flattening, results in a slowdown of propagating facilitated Rydberg excitations. This effect is shown in Fig.~\ref{Fig:Fig3}(a), where we display the real-time dynamics of both Rydberg excitations and phonons. For the simulations we performed exact diagonalization on a system of size $N=10$ and we truncated the local bosonic Hilbert space allowing a maximum number of three bosons per site. The initial state contains a single Rydberg excitation at the left edge of the lattice and no bosons, i.e. $|\uparrow\downarrow\downarrow\dots\rangle\bigotimes|0,0,0\dots\rangle$. Consequently, such wave packet states  of the form $|\psi_{\tn{in}}\rangle = |\dots\uparrow\downarrow\downarrow\dots\rangle$ in real space correspond to  superpositions of momentum states that live on the first two excited bands due to a mixing between the states introduced in the diagonalization of the Hamiltonian~\eqref{Eq:Ham_int}. 

The data in Fig.~\ref{Fig:Fig3}(a) shows that the stronger the coupling $\kappa$ the more pronounced becomes the phonon trail that is carried and left behind by the propagating Rydberg excitation. In Rydberg quantum simulator experiments it is standard to measure the Rydberg density \cite{Browaeys2020}. It is, however, also possible to determine the local phonon density by side-band spectroscopy, as demonstrated in Ref. \cite{Kaufman2012}. Remarkably, this makes it possible to use  Rydberg quantum simulators  to directly  detect and map out the phonon cloud in-situ and in real-time, which remains elusive in solid state systems and most ultracold atom platforms.

The magnitude of the phonon dressing can be quantified by the \emph{Z-factor} which  is defined by the overlap of the dressed polaron state $\ket{\bar\psi_q}$ with its non-interacting counterpart $\ket{\psi^{(0)}_q}$, $Z_q = |\langle\psi^{(0)}_q|\bar{\psi}_q\rangle|^2$~\cite{Alexandrov1996, landau1980statistical, Anderson1967}.  The calculation of the $Z$-factor from exact diagonalization, see Fig.~\ref{Fig:Fig3}(b) shows that, although the phonon dressing is strong, still a well-defined polaron quasiparticle exists.\\
We also compute the phonon occupation number in momentum space in the dressed ground state $|\bar{\psi}_q\rangle$, i.e. $n_{\tn{ph}}(p,q)$, with $p$ and $q$ being the phonon and quasi-particle momentum respectively, $n_{\tn{ph}}(p,q)=\langle\bar{\psi}_q|A_p^\dag A_p|\bar{\psi}_q\rangle$.
While this quantity cannot be computed exactly analytically, at  first non-zero order one finds:
\begin{eqnarray}
n^{(1)}_{\tn{ph}}(q,p)&=&\frac{\kappa^2}{\omega^2 N}\bra{\psi^{(0)}_q}\left(\e ^{\ii\frac{2\pi }{N}p}-1\right)\left(\e ^{-\ii\frac{ 2\pi}{N}p}-1\right)\ket{\psi^{(0)}_q}\nonumber\\
&=&2\frac{\kappa^2}{\omega^2 N}\left[1-\cos\left(\frac{2\pi}{N}p\right)\right],
\label{Eq:numbos1}
\end{eqnarray}
where $-\pi\leq \frac{2\pi}{N}p\leq \pi$. 
Note, that this result does not depend on the quasiparticle momentum $q$. Such a dependence enters at higher order in perturbation theory and  leads to a $q$-dependent coefficient to Eq.~\eqref{Eq:numbos1}. In fact our numerical calculations confirm a dependence of the form
\begin{equation}
n_{\tn{ph}}(p,q) = 2\frac{\kappa^2}{\omega^2 N}C^\ast(q)\left[1-\cos\left(\frac{2\pi}{N}p\right)\right], 
\end{equation}
with a numerically determined coefficient $C^\ast(q)$. In Fig. \ref{Fig:Fig3}(c) which we show the phonon occupation number $n_{\tn{ph}}(p,q)$ at the edges of the ground state band, $\frac{2\pi}{N}q=\pm\pi$, at $\kappa=0.3\Omega$. The agreement between numerical and analytical results from Eq. \eqref{Eq:numbos1} is excellent.

\emph{Conclusions.}-- 
We have shown how the non-equilibrium dynamics of a facilitated Rydberg atoms chain is dramatically affected by interactions with trap vibrations. This coupling leads to a dressing of the propagating excitations and shows the emergence of a slow-dynamics induced by a flattening of the quasi-particle bands. The latter can be interpreted as a polaronic effect that leads to an increase of the effective mass. The phonon dressing, as discussed here, might have links to other timely research questions: it was recently pointed out, that lattice Hamiltonians coupled to bosons can offer a possible setup for the observation of \emph{fractons} \cite{2019arXiv190408424S}, which are currently much studied in the context of ergodicity breaking in quantum systems. Moreover, tuning the interaction between the excitations and the phonons permits to control the spreading of information within the system, which is a timely theme in the domain of quantum technology~\cite{gate1, gate2}.

\acknowledgements \emph{Acknowledgements.}-- We acknowledge discussions with C. Gro\ss, W. Li and F. Gambetta. I.L. and R.S.  acknowledge support  from  the  DFG  through  SPP  1929  (GiRyd). I.L. is  acknowledging support by the  “Wissenschaftler-R\"uckkehrprogramm GSO/CZS”  of  the  Carl-Zeiss-Stiftung  and  the  German Scholars Organization e.V. R.~S. is supported by the Deutsche Forschungsgemeinschaft (DFG, German Research Foundation) under Germany's Excellence Strategy -- EXC-2111 -- project ID 390814868. 

\begin{widetext}

\newpage 
\begin{center}
\textbf{Supplemental material: Vibrational dressing in kinetically constrained Rydberg spin systems}
\end{center}

In this supplemental material we show step-by-step how the Hamiltonian (3) in the main text can be rewritten as in Eq.~(7). We will also write explicitly all the interaction terms.

\section{Hamiltonian in the effective space}
Let us start by considering the Hamiltonian describing Rydberg atoms in the effective ``constrained'' Hilbert space. This reads (see Eq.~(4) in the main text):
\begin{equation}
\begin{split}
H = &\Omega\Big(\sum_{\alpha}|\alpha\rangle\langle \alpha| \mu^x +\mu^-|\alpha+1\rangle\langle\alpha| +\tn{h.c.}\Big) \\
&+\kappa\sum_{\alpha}\frac{\mu^z-\mathbb{1}}{2}|\alpha\rangle\langle \alpha| (a^\dag_{\alpha+1} + a_{\alpha+1} - a^\dag_\alpha - a_\alpha)\\
&+ \omega\sum_{\alpha} a^\dag_\alpha a_\alpha,
\end{split}
\label{Eq:Hamiltonian}
\end{equation}
 where the $\mu$-operators are the ones defined in the main text. The first step is to move to the Fourier space for the bosonic modes of the harmonic traps. This is achieved by defining
\begin{equation}
a_m=\frac{1}{\sqrt{N}}\sum_{p=-N/2}^{N/2}A_p \e^{\frac{\ii 2\pi}{N}mp}.
\end{equation}
We thus see that the difference between the phonon creation operators appearing in the interaction term can be rewritten as
\begin{equation}
a^\dag_{m+1}-a^\dag_m=\frac{1}{\sqrt{N}}\sum_p\left[\left(\e^{-\frac{\ii(m+1)2\pi}{N}p}-\e^{-\frac{\ii m 2\pi}{N}p}\right)A^\dag_x\right].
\end{equation}
As we showed in the main text (see Eq.~(4))) this leads to the Hamiltonian
\begin{equation}
\begin{split}
H = &\Omega\sum_{\alpha}|\alpha\rangle\langle\alpha|\mu^x + \Omega\sum_\alpha\big(\mu^-|\alpha +1\rangle\langle\alpha| + \tn{h.c.}\big)\\
&+\frac{\kappa(\mu^z-\mathbb{1})}{2\sqrt{N}}\sum_p \left[\left(\e ^{-\ii\frac{ 2\pi}{N}p}-1\right)\e^{-\ii\frac{ 2\pi}{N}\hat{\alpha}}A^\dag_p+\textnormal{h.c.}\right]\\
&+\omega\sum_p A^\dag_pA_p,
\end{split}
\label{Eq:Hamiltonian1}
\end{equation}
 in which $\hat{\alpha}=\sum_\alpha\alpha\ket{\alpha}\bra{\alpha}$. At this point we can apply the Lee-Low-Pines transformation, which is defined as
\begin{align}
U = &\exp{\left[-\ii\hat \alpha  \sum_p \frac{2\pi p}{N} A^\dag_pA_p\right]}\\
U^\dag = &\exp{\left[\ii\hat \alpha  \sum_p \frac{2\pi p}{N} A^\dag_pA_p\right]}.
\label{Eq:Lee-Low-Pines}
\end{align}
We stress, again, that this transformation is important because it decouples the lattice degrees of freedoms from the phonons. Applying the transformation \eqref{Eq:Lee-Low-Pines} to the operators in Eq.~\eqref{Eq:Hamiltonian1} we have:
\be
U^\dag A_p U=\exp\left\{-\ii \frac{2\pi p}{N}\hat{\alpha}\right\}A_p,
\ee
and
\be
U^\dag |m+1\rangle\langle m|U=\e^{+\ii\sum_pA^\dag_pA_p\frac{2\pi p(m+1)}{N}}|m+1\rangle\langle m|\e^{-\ii\sum_pA^\dag_pA_p\frac{2\pi p m}{N}}=|m+1\rangle\langle m|\e^{\frac{-\ii2\pi}{N} p\sum_p A^\dag_p A_p}.
\ee
Therefore, Hamiltonian~\eqref{Eq:Hamiltonian1} can be rewritten as
\begin{multline}
U^\dag \HH U= \Omega\sum_\alpha \left[\aal\mu^x+\aaa \e^{-\ii\frac{2\pi}{N}p\sum_pA^\dag_pA_p}\mu^-+\aaadag \e^{\ii\frac{2\pi}{N}\sum_p pA^\dag_pA_p}\mu^+\right]\\
 +\frac{\kappa(\mu^z-\mathbb{1})}{2\sqrt{N}}\sum_x\left[\left(\e^{-\ii\frac{2\pi}{N}p}-1\right)\e^{-\ii\frac{ 2\pi}{N}\hat\alpha}A^\dag_p+\tn{h.c.}\right]+\omega\sum_p A^\dag_pA_p.
\end{multline}
In order to get the rid of the lattice labels $\alpha$ we move to the Fourier space for the quasi-particles:
\be
|\alpha\rangle=\frac{1}{\sqrt{N}}\sum_{q=-N/2}^{N/2} \e^{\frac{\ii\alpha2\pi q}{N}}|q\rangle
\ee
We then obtain
\begin{multline}
\HH=\Omega\sum_q|q\rangle\langle q|\left[\mu^x+\mu^-\e^{-\ii\frac{2\pi(\sum_p pA^\dag_pA_p+q)}{N}}+\mu^+\e^{+\ii\frac{2\pi(\sum_p pA^\dag_pA_p+q)}{N}}\right]+\\
\frac{\kappa(\mu^z-\mathbb{1})}{2\sqrt{N}}\sum_x\left[\left(\e^{-\ii\frac{2\pi}{N}p}-1\right)A^\dag_p+\tn{h.c.}\right]+\omega\sum_p A^\dag_pA_p.
\label{Eq:Hamiltonian2}
\end{multline}
Note, that Hamiltonian~\eqref{Eq:Hamiltonian2} is diagonal in the quasi-particles momentum $q$. Hence, we can diagonalize for every $q$ the free part of it, i.e. the Hamiltonian corresponding to $\kappa=0$.
\section{Diagonalization of the free part}
Let us rewrite Eq.~\eqref{Eq:Hamiltonian2} in matrix form, i.e. writing explicitly the matrices $\mu^{x,\pm}$ and completing the squares for the bosonic part 
\begin{multline}
\HH_q=\Omega \begin{pmatrix}
0 & \e^{+\ii\frac{2\pi(\sum_p pA^\dag_pA_p+q)}{N}}+1 \\
\e^{-\ii\frac{2\pi(\sum_p pA^\dag_pA_p+q)}{N}}+1 & 0
\end{pmatrix}+\\
+\omega\sum_p\left[A_p+\frac{\kappa}{\omega\sqrt{N}}\left(\e^{+\ii\frac{2\pi}{N}p}-1\right)\begin{pmatrix}
1 & 0 \\
0 & 0
\end{pmatrix}\right]^\dag\left[A_p+\frac{\kappa}{\omega\sqrt{N}}\left(\e^{-\ii\frac{2\pi}{N}p}-1\right)\begin{pmatrix}
1 & 0 \\
0 & 0
\end{pmatrix}\right]+\\
-\frac{\kappa^2}{\omega N}\sum_p 2\left[1-\cos\left(\frac{2\pi}{N}p\right)\begin{pmatrix}
1 & 0 \\
0 & 0
\end{pmatrix}\right].
\label{Eq:Ham_matrix}
\end{multline}
Defining a \emph{displacement} operator for the bosons, i.e.
\be
\D=\exp\left[-\frac{\kappa}{\sqrt{N}\omega}\sum_p\left(\e^{-\ii\frac{2\pi}{N}p}-1\right)A^\dag_p-\tn{h.c.}\right]
\label{Eq:displacement}
\ee
such that $\D^\dag \At_p \D = A_p$, with $\tilde{A}_p= A_p +\frac{\kappa}{\omega\sqrt{N}}\left(\e ^{-\ii\frac{2\pi }{N}p}-1\right)$, we can cast Eq.~\eqref{Eq:Ham_matrix} in the following form:
\begin{equation}
\D^\dag\HH_q\D=\omega\sum_p \At^\dag_p\At_p+\Omega\begin{pmatrix}
0 & \e^{+\ii\frac{2\pi(\sum_p p A ^\dag_p A_p+q)}{N}}+1 \\
\e^{-\ii\frac{2\pi(\sum_p p A^\dag_p A_p+q)}{N}}+1 & 0
\end{pmatrix}-2\frac{\kappa^2}{\omega}\nn+\tilde{\hat{H}}^{\tn{II}}_{\tn{int}}
\label{Eq:Ham_matrix1}
\end{equation}
Note, that the effect of the interaction between the lattice and the phonons is only in the argument of the displacement operator. We can now diagonalize the off-diagonal matrix appearing in~\eqref{Eq:Ham_matrix1}. 
Casting $\theta=\sum_p\left(p A_p^\dag A_p +q\right)$, the matrix we want to diagonalize has therefore the form
\be
\begin{pmatrix}
0 & \e^{2 \ii \theta\pi/N} +1\\
\e^{-2 \ii \theta\pi/N}+1 & 0
\end{pmatrix}.
\ee 
Its eigenvectors are $\begin{pmatrix}  -\e^{\ii \theta\pi/N}\\1 \end{pmatrix}$ and $\begin{pmatrix}  \e^{\ii \theta\pi/N}\\1 \end{pmatrix}$, therefore the unitary matrix $S$ which implements the diagonalization is
\be
S=\begin{pmatrix}  
-\e^{\ii \theta\pi/N} & \e^{\ii \theta\pi/N}\\
1 & 1
\end{pmatrix}.
\ee
The diagonalization induces a mixing between the states $|q,\mu^z=1\rangle$ and $|q, \mu^z=2\rangle$.\\
The term $\hat{H}^{\tn{II}}_{\tn{int}} = S^\dag \tilde{\hat{H}}^{\tn{II}}_{\tn{int}} S$ is obtained by the action of the displacement operator $\hat{D}$, definined in Eq~\eqref{Eq:displacement}, on the Rabi part of the Hamiltonian Eq.~\eqref{Eq:Ham_matrix}. We want to derive an effective expression for this interaction term in the perturbative limit. In the limit of small $\kappa$ the can rewrite the displacement operator as
\begin{equation}
\hat{D} = \e^{-\kappa \sum_{p}\alpha_p \hat{A}^\dag_p + \kappa\sum_p\alpha_p^\ast\hat{A}_p}\simeq\left[\mathbb{1} - \kappa\sum_p(\alpha_p \hat{A}^\dag_p - \alpha_p^\ast\hat{A}_p) +\frac{\kappa^2}{2}\left(\sum_p(\alpha_p^\ast \hat{A}_p - \alpha_p\hat{A}_p^\dag)\right)^2\right]+\dots,
\end{equation}
where $\alpha_p=\frac{1}{\omega\sqrt{N}}\left(\e^{-\ii\frac{2\pi}{N}p}-1\right)$. Therefore, we have
\begin{equation}
\begin{split}
&\hat{D}^\dag H_f\hat{D} \simeq\\ 
&\left[\mathbb{1} - \kappa\sum_p(-\alpha_p \hat{A}^\dag_p + \alpha_p^\ast\hat{A}_p) +\frac{\kappa^2}{2}\left(\sum_p(-\alpha_p^\ast \hat{A}_p + \alpha_p\hat{A}_p^\dag)\right)^2\right] H_f \left[\mathbb{1} - \kappa\sum_p(\alpha_p \hat{A}^\dag_p - \alpha_p^\ast\hat{A}_p) +\frac{\kappa^2}{2}\left(\sum_p(\alpha_p^\ast \hat{A}_p - \alpha_p\hat{A}_p^\dag)\right)^2\right]. 
\end{split}
\end{equation}
From which we obtain, order by order in $\kappa$
\begin{equation}
\begin{split}
H_f + V =
&H_f - \kappa\left[\sum_p(-\alpha_p \hat{A}^\dag_p + \alpha_p^\ast\hat{A}_p)H_f + H_f\sum_p(\alpha_p \hat{A}^\dag_p - \alpha_p^\ast\hat{A}_p)\right] + \\
&\kappa^2\left[\sum_p(\alpha_p \hat{A}^\dag_p - \alpha_p^\ast\hat{A}_p)H_f\sum_p(-\alpha_p \hat{A}^\dag_p + \alpha_p^\ast\hat{A}_p)\right]+\\
&\frac{\kappa^2}{2}\left[\left(\sum_p(\alpha_p \hat{A}^\dag_p - \alpha_p^\ast\hat{A}_p)\right)^2H_f + H_f\left(\sum_p(-\alpha_p \hat{A}^\dag_p + \alpha_p^\ast\hat{A}_p)\right)^2\right]+\dots
\end{split}
\label{Eq:int_iii}
\end{equation}
At this point we can diagonalize $H_f$ in Eq.~\eqref{Eq:int_iii} obtaining
\begin{equation}
-\Omega\cos \left[ \frac{\pi}{N}\left(\sum_pp A^\dag_pA _p+q\right)\right]\mu^z + S^\dag V S.
\end{equation}
The interaction term $S^\dag V S$ is quite complicated, however as long as we are interested in the first order correction on the ground state, we have that
\begin{equation}
H^{\tn{II}}_{\tn{int, GS}}(q) = \frac{\Omega\kappa^2}{\omega^2 N} \sum_p |\alpha_p|^2 A_p\cos\left(p A^\dag_pA_p + q\right) A^\dag_p.
\end{equation}
This term contributes to the energy correction reported in the main text.\\
The complete Hamiltonian is therefore
\begin{equation}
    H_q = \omega\sum_p \tilde{A}^\dag_p\tilde{A}_p\mathbb{1}-\Omega \cos\left[\frac{\pi}{N}\left(\sum_p p A^\dag_p A_p+q\right)\right]\mu^z
     +H^{\tn{I}}_{\tn{int}} + H^{\tn{II}}_{\tn{int}},
\end{equation}
with
\begin{equation}
    H^{\tn{I}}_{\tn{int}} = -\frac{\kappa^2}{\omega}(\mathbb{1}-\mu^x)
\end{equation}
and,
\begin{equation}
    H^{\tn{II}}_{\tn{int}} = S^\dag V S
\end{equation}
For the leading order correction to the energy of the ground state band $E_{\tn{GS}}(q)$ the only terms which gives a non-zero contribution are $H^{\tn{I}}_{\tn{int}}$ and the $H^{\tn{II}}_{\tn{int, GS}}$ (see discussion in main text).

\section{Expansion of the potential}
In this section we justify the approximation reported in Eq. (2) in the main text. Let us consider a generic potential of a one-dimensional lattice embedded in two dimensions. This means that we can have fluctuations around the equilibrium position in two directions that we will call $z$ for the longitudinal one and $y$ for the transverse one. Without loss of generality we can suppose that the interaction depends only on the relative distance between two atoms, i.e.
\begin{equation}
V(\mathbf{r}_i, \mathbf{r}_j) = V (|\mathbf{r}_i-\mathbf{r}_j|) = V (|r_{i,z} - r_{j,z}|, |r_{i,y} - r_{j,y}|) = V (r_z, r_y)
\end{equation}
where $r_z = r_{i,z} - r_{j,z} $ and $ r_y = r_{i,y} - r_{j,y} $. For one-dimensional lattices, considering only nearest-neighbours interaction, the equilibrium positions of the atoms are $r_z = a$, with $a$ the lattice spacing and $r_y = 0$. Performing the expansion we obtain
\begin{equation}
V(r_z, r_y) = V(a, 0) + \left. \frac{\partial V(r_z, r_y)}{\partial r_z} \right|_{r_z = a}\delta r_z + \left. \frac{\partial V(r_z, r_y)}{\partial r_y} \right|_{r_y = 0}\delta r_y+\dots
\end{equation}
As we reported in the main text, we can rewrite the displacement $\delta r_{z,y}$ in terms of the bosonic operators, the coupling is proportional to oscillator length, i.e.
\begin{equation}
\delta r_{\mu} = l_\mu(a^\dag_i + a_i -a^\dag_j- a_j) \qquad \mu = z,y.
\end{equation} 
Here $l_\mu = \sqrt{\hbar/(m\omega_\mu)}$ is the harmonic oscillator length. It is possible to observe how tuning the trapping frequency in a different way in the two directions leads to a different coupling with the transverse and longitudinal modes. For the general case of a power-law decaying potential we have that:
\begin{equation}
V(r_z, r_y)\propto \frac{1}{(r_z^2 + r_y^2)^{\frac{\alpha}{2}}},
\end{equation}
therefore,
\begin{equation}
\left. \frac{\partial V(r_z, r_y)}{\partial r_y} \right|_{r_y = 0} = \left.\frac{\alpha r_y}{(r_z^2 + r_y^2)^{\frac{\alpha}{2}+1}}\right |_{r_y=0}=0.
\end{equation}
This shows that, at first order, the contribution of the transverse modes to the longitudinal interaction is zero.

\section{Experimental considerations}
In this section we give some remarks concerning the parameters of a possible experimental realisation of the system. We focus here on $^{87}$Rb and $^{133}$Cs. However, the order of magnitude of the parameters is comparable to that of other experiment conducted e.g. with $^{39}$K and $^7$Li. We will also explain more in detail how the observables discussed in the paper can be detected in an experiment. \\
Let us start by giving some typical values for the trap parameters that are usually set in optical tweezers experiments. The lattice constant $a$, i.e. the distance between the Rydberg atoms, is $a\approx 5\mu$m. The life-time of the Rydberg state with high principal quantum number $n$, $n\simeq 40-50$, is approximately $\tau_R\simeq 2\cdot10^{-5}$s. The trapping frequency $\omega$ is typically $\omega \simeq 2\pi\cdot 300$kHz, the Rabi frequency $\Omega$ can be tuned until a maximum value of $\Omega_{max} \simeq 2\pi\cdot 10$MHz. The Van der Walls constant between  $ns$-states scales with the Rydberg principal number $n$ as $C_6([ns]) = n^{11}(c_0+c_1n + c_2n^2)$au~\cite{Singer_2005}. For $^{87}$Rb we have $c_0 = 11.97$, $c_1 = -0.8486$ and $c_3 = 3.385\cdot10^{-3}$, for  $^{133}$Cs, instead, $c_0 = 10.64$, $c_1 = -0.6249$ and $c_3 = 2.33\cdot10^{-3}$. This leads, for $n=43$, to an interaction strength between nearest neighbours of $V_{int, \textnormal{Rb}}\approx 1\cdot$MHz and $V_{int, \textnormal{Cs}}\approx 0.65\cdot$MHz. In the case studied in this paper, what matters is not the interaction in itself (since we are in the facilitation regime) but its gradient, i.e. $G = -6\frac{V_{int}}{a}$. Considering the same parameters as before we obtain that $G_{\textnormal{Rb}} = -1.2\cdot10^3$kHz$\mu$m$^{-1}$ and $G_{\textnormal{Cs}} = -7.8\cdot10^2$kHz$\mu$m$^{-1}$ . The interaction constant $\kappa$ is related to the gradient via the harmonic oscillator length, i.e. $\kappa = -\frac{l_{ho}}{\sqrt{2}}G$. With the previous data, we have: $\kappa_{\textnormal{Rb}}\approx 16$kHz and $\kappa_{\textnormal{Cs}}\approx 9$kHz.  Experimentally, these coupling constants can be controlled using microwave-dressing of Rydberg $s-$ and $p-$states, as discussed in Ref.~\cite{Gambetta}. This procedure enable us to tune independently the gradient from the interaction.\\
The many-body dynamics can be characterised by measuring the spin (Rydberg) density and the phonon density, as shown in Fig.~3 in the main text. The spin density can be detected by counting the atoms in the Rydberg state, this can be achieved using projective measurements (see for example \cite{Browaeys2020}). However, in these experiments we can also detect the phonon density, which is particularly interesting because it enables us to measure directly the effect of the dressing of the excitations. This can be done using side-band spectroscopy (as shown in Ref.~\cite{Kaufman2012}). As stated in the main text, the combination of the detection methods and the exaggerated length scales offer unique opportunities for investigating polaron physics. 

\end{widetext}
\bibliography{bib}

\end{document}